\documentstyle[emulateapj]{article}
\def\vx{{\bf x}}

\def\omb{{\Omega_{\mathrm b}}}
\def\vp{{v_\parallel}}
\begin{document}
\submitted{ApJ Letters, accepted, March 2, 2000}
\title{Large scale motions in superclusters: their imprint in the CMB}
\author{Antonaldo Diaferio\altaffilmark{1,2}, Rashid A. Sunyaev\altaffilmark{1,3}, and
Adi Nusser\altaffilmark{4}}
\altaffiltext{1}{Max-Planck Institut f\"ur Astrophysik, Karl-Schwarzschild-Str. 1,
D-85740 Garching, Germany}
\altaffiltext{2} {Present address: Universit\`a di Torino, Dipartimento
di Fisica Generale ``Amedeo Avogadro'', Via P. Giuria 1, I-10125 Torino, Italy}
\altaffiltext{3}{Space Research Institute (IKI), Profsouznaya 84/32, Moscow 117810, Russia}
\altaffiltext{4}{Physics Department, Technion-Israel Institute of Technology, Technion City,
Haifa 32000, Israel}
\begin{abstract}

We identify high density regions of supercluster size
in high resolution $N$-body simulations of
a representative volume of three Cold Dark Matter Universes. 
By assuming that (1) the density and peculiar velocities
of baryons trace those of the dark matter, and (2) 
the temperature of plasma is proportional to 
the velocity dispersion of the dark matter particles 
in regions where the crossing
times is smaller than the supercluster free-fall time, we investigate
how thermal motions of electrons in the intra-cluster
medium  and peculiar velocity of clusters can 
affect the secondary anisotropies in the cosmic microwave background (CMB).
We show that the thermal
effect dominates the kinematic effect and that
the largest thermal decrements are associated with the most
massive clusters in superclusters. Thus, searching for the presence of two or more close large 
CMB decrements represents a viable strategy for identifying
superclusters at cosmological distances.
Moreover, maps of the kinematic effect in
superclusters are characterized by neighboring large peaks of opposite
signs. These peaks can be as high as $\sim 10\mu$K at
the arcminute angular resolution.  Simultaneous pointed observations
of superclusters in the millimeter and submillimeter bands with upcoming sensitive CMB
experiments can separate between the thermal and kinematic effect
contributions and constrain the evolution of the velocity field in
large overdense regions.
\end{abstract}


\section{INTRODUCTION}

Thermal motions of ionized gas in the intra-cluster medium and the
bulk flow of clusters produce secondary temperature fluctuations in
the cosmic microwave background (CMB) on the arcminute scale
(\cite{sunyaev70}, 1972). For a given cluster velocity and mass, these
thermal and kinematic effects (tSZ and kSZ effects,
hereafter) are independent of redshift.  
Typical line-of-sight peculiar velocities of galaxy clusters,
estimated from distance indicators to cluster galaxy members, are
$\sim 300$ km s$^{-1}$ (e.g. \cite{giov98}).  
These velocities can easily be $\sim 50\%$ larger
for clusters in supercluster environments (\cite{colberg99}).
Therefore, CMB anisotropies caused by the kSZ effect in superclusters
will be substantially enhanced.  

Ideally, one would like to estimate
the magnitude of SZ anisotropies  using hydro-dynamical simulations.
Unfortunately, these simulations do not currently have the necessary
dynamical range to study, at the same time, 
the large scale motions of clusters and the
dynamics of the gas inside them. Therefore, we resort to high
resolution $N$-body simulations of dark matter particles only. 
By assuming that the motion and distribution of electron gas 
follow the dark matter, these simulations can be used to estimate the
SZ anisotropies in the CMB.
Here, we show that superclusters have a clear signature
in both tSZ and kSZ maps: in tSZ maps, superclusters
appear as regions of neighboring large CMB decrements; in kSZ
maps, supercluster regions contain large neighboring peaks of 
opposite sign. The first property can provide an easy
strategy for indentifying superclusters at any redshift; after
removing the dominating tSZ effect contribution, the detectable kSZ peaks
provide a direct measure of the velocity field in superclusters.

\section{Method}
\label{method}

To explore the non-linear evolution of  the 
density and velocity field of superclusters, we use
the $N$-body simulations of three variants of the
Cold Dark Matter (CDM) model 
(see \cite{kauffmann99} for details): two low
density models with a
$141^3 h^{-3}$ Mpc$^3$ comoving box\footnote{We use
$H_0=100\ h\ {\rm km\ s}^{-1}\ {\rm Mpc}^{-1}$ throughout.}
 (OCDM with $\Omega_0=0.3$; $\Lambda$CDM 
with $\Omega_0=0.3$, $\Lambda=0.7$) and 
a high density model with a $85^3 h^{-3}$ Mpc$^3$ comoving box 
($\tau$CDM with $\Omega_0=1$). 
All models contain $256^3$ particles and 
are normalized to the present abundance of galaxy clusters.

In order to locate superclusters in the simulation, we compute the 
density $\rho({\vx})$ on a $256^3$ grid from the particle positions at $z=0$
with the cloud-in-cell (CIC) interpolation scheme.
We then smooth the density field with a  gaussian  window
of width $4h^{-1}$ Mpc,  compute the
average overdensity profiles $\delta(<r)$ 
around local maxima and search for 
regions with $\delta(r<8 h^{-1} {\rm Mpc})\equiv\delta_8\ge \delta_s$,
where $\delta_s$ is a parameter. The chosen window width
identifies superclusters efficiently: smaller
widths also identify isolated clusters; larger widths miss
low density superclusters.
Because we set the same initial random phases in all our simulations, they
yield the same large scale matter distribution at $z=0$. For example,
in the OCDM model at $z=0$, we find 45, 15, 7, 3, and 1 regions
with $\delta_s=2$, 3, 4, 5, and 6, respectively.  
The most dense region has $\delta_8=6.8$.
Hereafter, we define as superclusters the regions with $\delta_8\ge 4$ 
(e.g. \cite{einasto97}). These superclusters have crossing
times larger than the Hubble time and they are not in dynamical
equilibrium. Clusters within the simulation
box are identified with a friends-of-friends algorithm applied
to the dark matter particles. 

\begin{table*}
{\scriptsize
\begin{center}
\centerline{\sc Table 1}
\vspace{0.1cm}
\centerline{\sc Evolution of the Properties of Supercluster Regions}
\vspace{0.3cm}
\begin{tabular}{ccccccccccc}
\hline\hline
\noalign{\smallskip}
\multicolumn{1}{c}{\#} & 
\multicolumn{1}{c}{$z$} & 
\multicolumn{1}{c}{$\tau_T$} &
\multicolumn{1}{c}{$\langle\sigma_\parallel^2\rangle^{1/2}$ } &
\multicolumn{1}{c}{$\langle\vp\rangle$ } & 
\multicolumn{1}{c}{tSZ } & 
\multicolumn{1}{c}{kSZ } & 
\multicolumn{1}{c}{rms $\tau_T$ } &
\multicolumn{1}{c}{rms tSZ }  &
\multicolumn{1}{c}{rms kSZ} & 
\multicolumn{1}{c}{$\delta r$} \cr 
 & &$\times 10^{-3}$ & km s$^{-1}$ & km s$^{-1}$ & $\mu$K & $\mu$K  & $\times 10^{-3}$ & $\mu$K & $\mu$K & $h^{-1}$ Mpc \cr
(1) & (2) & (3) & (4) & (5)  &  (6)  & (7)  & (8) & (9) & (10) &(11) \cr
\hline
\noalign{\medskip}

  1&$  0.0$&$  (9.4,10.2)$&$  (1052,968)$&$   (138,-297)$&$  (-205,-193)$&$     (4,-10)$&$   0.5/   0.1$&$   9.4/   0.4$&$   0.6/   0.3$&$   3.38$\cr
  1&$  0.3$&$ (10.4,6.8)$&$  (1206,1270)$&$   (447,-810)$&$  (-287,-217)$&$    (17,-20)$&$   0.7/   0.4$&$  15.0/   3.7$&$   1.1/   0.7$&$   0.34$\cr
  1&$  0.5$&$  (9.8,10.6)$&$ (1201,902)$&$   (361,449)$&$  (-284,-171)$&$ (12,16)$&$   0.7/   0.4$&$  12.3/   3.4$&$   0.9/   0.6$&$   0.72$\cr
  1&$  0.7$&$ (14,6.5)$&$   (983,1003)$&$   (-12,-159)$&$  (-277,-134)$&$     (0,-4)$&$   0.7/   0.5$&$  10.0/   4.1$&$   0.9/   0.4$&$   4.28$\cr
  1&$  1.1$&$ (13.2,10.4)$&$ (744,722)$&$(106,12)$&$(-149,-110)$&$ (5,1)$&$   0.8/   0.5$&$   6.1/   2.3$&$   0.8/   0.6$&$   3.45$\cr
\cr
  2&$  0.0$&$ (8.7,3.9)$&$(997,617)$&$(-84,-254)$&$ (-172,-29)$&$(-3,-3)$&$   0.4/   0.2$&$   5.9/   1.4$&$   0.4/   0.3$&$   1.48$\cr
  2&$  0.3$&$(10.1,5.4)$&$ (948,551)$&$(-11,-55)$&$(-185,-32)$&$( 0,-1)$&$   0.5/   0.3$&$   6.2/   0.8$&$   0.5/   0.5$&$   4.48$\cr
  2&$  0.5$&$ (10.9,8)$&$ (969,722)$&$(-577,174)$&$(-206,-85)$&$(-23,5)$&$   0.5/   0.4$&$   6.8/   1.4$&$   0.8/   0.5$&$   0.90$\cr
  2&$  0.7$&$ (9.8,9.3)$&$  (745,753)$&$(-515,9)$&$(-111,-106)$&$(-17,0)$&$   0.5/   0.3$&$   4.6/   0.7$&$   0.7/   0.4$&$   2.20$\cr
  2&$  1.1$&$(11.6,9.5)$&$(749,678)$&$(-338,-15)$&$(-131,-89)$&$(-12,0)$&$   0.6/   0.4$&$   4.9/   0.7$&$   0.6/   0.4$&$   3.27$\cr
\cr
  5&$  0.0$&$(6.3,5.2)$&$ (789,718)$&$  (-367,568)$&$(-80,-55)$&$(-8,10)$&$   0.3/   0.1$&$   2.8/   0.1$&$   0.4/   0.1$&$   1.36$\cr
  5&$  0.3$&$(6.9,7.3)$&$(742,706)$&$(-350,408)$&$(-77,-75)$&$(-8,10)$&$   0.4/   0.1$&$   3.1/   0.1$&$   0.4/   0.1$&$   2.03$\cr
  5&$  0.5$&$ (8.4,7.8)$&$  (710,703)$&$(430,-346)$&$(-86,-78)$&$(12,-9)$&$   0.4/   0.2$&$   3.3/   1.0$&$   0.5/   0.2$&$   1.82$\cr
  5&$  0.7$&$ (8.8,6.9)$&$ (719,613)$&$(462,-327)$&$ (-91,-53)$&$ (14,-8)$&$   0.5/   0.3$&$   3.4/   1.5$&$   0.6/   0.2$&$   1.45$\cr
  5&$  1.1$&$ (10.9,7.3)$&$   (666,579)$&$   (427,-109)$&$  (-101,-49)$&$ (17,-3)$&$    0.6/   0.3$&$   3.4/   1.0$&$   0.7/   0.4$&$   4.20$\cr
\hline
\noalign{\smallskip}
\noalign{\hrule}
\noalign{\smallskip}
\label{tab:sc}
\end{tabular}

{Evolution of the properties of $8\times 8 h^{-2}$ Mpc$^2$ supercluster regions 
centered on the supercluster
positions at $z=0$. Values are from maps of the OCDM model with 
a comoving spatial resolution of $0.2 h^{-1}$ Mpc. Column (1) supercluster 
identification number in order of decreasing overdensity; column
(2) redshift; columns (3)--(7) optical depth, velocity dispersion, peculiar velocity,
tSZ and kSZ fluctuations of the largest (left entries in 
brackets) tSZ decrement and 
 second largest (right entries)  tSZ decrement; 
columns (8)--(10) rms values including/excluding 
two squares of $2h^{-1}$ Mpc on a side each centered on one of the two largest tSZ decrements; 
(11) projected comoving separation between the two largest tSZ decrements.}
\end{center}
}
\end{table*}

The tSZ and kSZ effects depend on the temperature and line of sight velocity
of the intra-cluster medium, and on the number density of free electrons.
Therefore we have to relate these quantities to the positions
and velocities of the dark matter particles in the simulations.
We first compute the density and velocity fields in the simulation box 
on a $2048^3$ grid with the CIC interpolation 
scheme. This grid 
preserves the $30 h^{-1}{\mathrm kpc}$ spatial
resolution of our simulations in the high density regions
we are interested in, namely
superclusters, filaments, clusters and groups of galaxies. 
A coarser grid would artificially lower the magnitude of 
the SZ fluctuations but not their relative location and sign.
We then assume that at any point $\vx$ within
the cluster the plasma is thermalized: 
$kT_e(\vx)=\sigma^2_{\parallel}(\vx) m_{\mathrm p}/2$, 
where $T_e$ is the gas temperature, $\sigma^2_{\parallel}$ the local 
one dimensional velocity dispersion of the dark matter particles, 
$m_{\mathrm p}$ the proton mass, and $k$ the Boltzmann constant.   
We also assume that the electron
number density, $n_{\mathrm e} ({\vx })$, is related to the mass
density $\rho({\vx})$ by $n_{\mathrm e}({\bf
x})=[\rho({\vx})/m_{\mathrm p}] \omb/\Omega_0$, where 
$\omb$ 
is the background baryonic 
density parameters in terms of the critical density. 
All calculations in this {\em Letter} are done
with $\omb=0.0125h^{-2}$ (e.g. \cite{smith93}).  Note however, that
clusters may contain more baryons than the average baryon fraction in the
Universe (e.g. \cite{fuku98}), so we might be underestimating the actual magnitude of the
SZ effects by a factor of a few.
 
Under these assumptions, the hot electron gas causes  
tSZ temperature decrements in the CMB spectrum
\begin{equation}
\left.{\Delta T\over T}\right\vert_{th} = \left(x{e^x+1\over e^x-1}-4\right)
{m_{\mathrm p}\over m_{\mathrm e}}
{\omb\over\Omega_0}\sigma_T\int_0^{L} {\sigma_\parallel^2(l)\over 2 c^2}
{\rho(l)\over m_{\mathrm p}} {\mathrm d}l
\label{eq:tsz}
\end{equation}
where $x=h\nu/k T_r$, $T_r$ is the radiation temperature, 
$L$ the simulation box proper size, $\sigma_T$ the Thomson cross-section,
$m_{\mathrm e}$ the electron mass, and $c$ the speed of light.  
Numerical values of the 
tSZ decrement presented here are computed in the Rayleigh-Jeans limit $x\to 0$.
Line-of-sight peculiar velocities $\vp$ produce kSZ temperature fluctuations
\begin{equation}
\left.{\Delta T\over T}\right\vert_{kin}=
{\omb\over\Omega_0}\sigma_T\int_0^{L} {\vp (l)\over c}
{\rho(l)\over m_{\mathrm p}} {\mathrm d}l.
\label{eq:ksz}
\end{equation}
With broad band observations, one can separate the thermal and kinematic 
effects
and use the latter to measure the peculiar velocity of clusters (\cite{sunyaev80}; 
\cite{rephaeli91}; \cite{haehnelt96}; \cite{aghanim97}). 

We obtain two dimensional maps by projecting the simulation box along a random line
of sight.  Hereafter, we present results for all models projected
along the same direction. 
We also compute maps of the Thomson optical depth $\tau_T =
(\omb/\Omega_0)\sigma_T\int_0^{L} {\rho(l)/ m_{\mathrm p}} {\mathrm d}l$, 
the line of sight velocity
dispersion $\langle \sigma_\parallel^2 \rangle=\int_0^{L}
\sigma_\parallel^2(l) \rho(l) {\mathrm d}l/\int_0^{L}\rho(l)
{\mathrm d}l$, and the line of sight velocity
field $\langle \vp\rangle=\int_0^{L} \vp(l)\rho(l) {\mathrm
d}l/\int_0^{L}\rho(l) {\mathrm d}l$.  

\section{results}
\label{results}

Figure \ref{fig:maptSZ} shows the tSZ decrements in the OCDM box at $z=0$.
The largest decrements are in high density regions: superclusters
appear as neighboring large decrements. This signature is characteristic
of supercluster regions. Because the tSZ effect is independent of redshift,
and, on arcminute angular scales, the tSZ decrements of the 
most massive clusters in a supercluster are
larger than the cosmological fluctuations,
neighboring large CMB decrements clearly  indicate the presence
of superclusters at cosmological distances.

\begin{figure*}
\epsscale{0.7}
\vskip -0.3 truein
\plotone{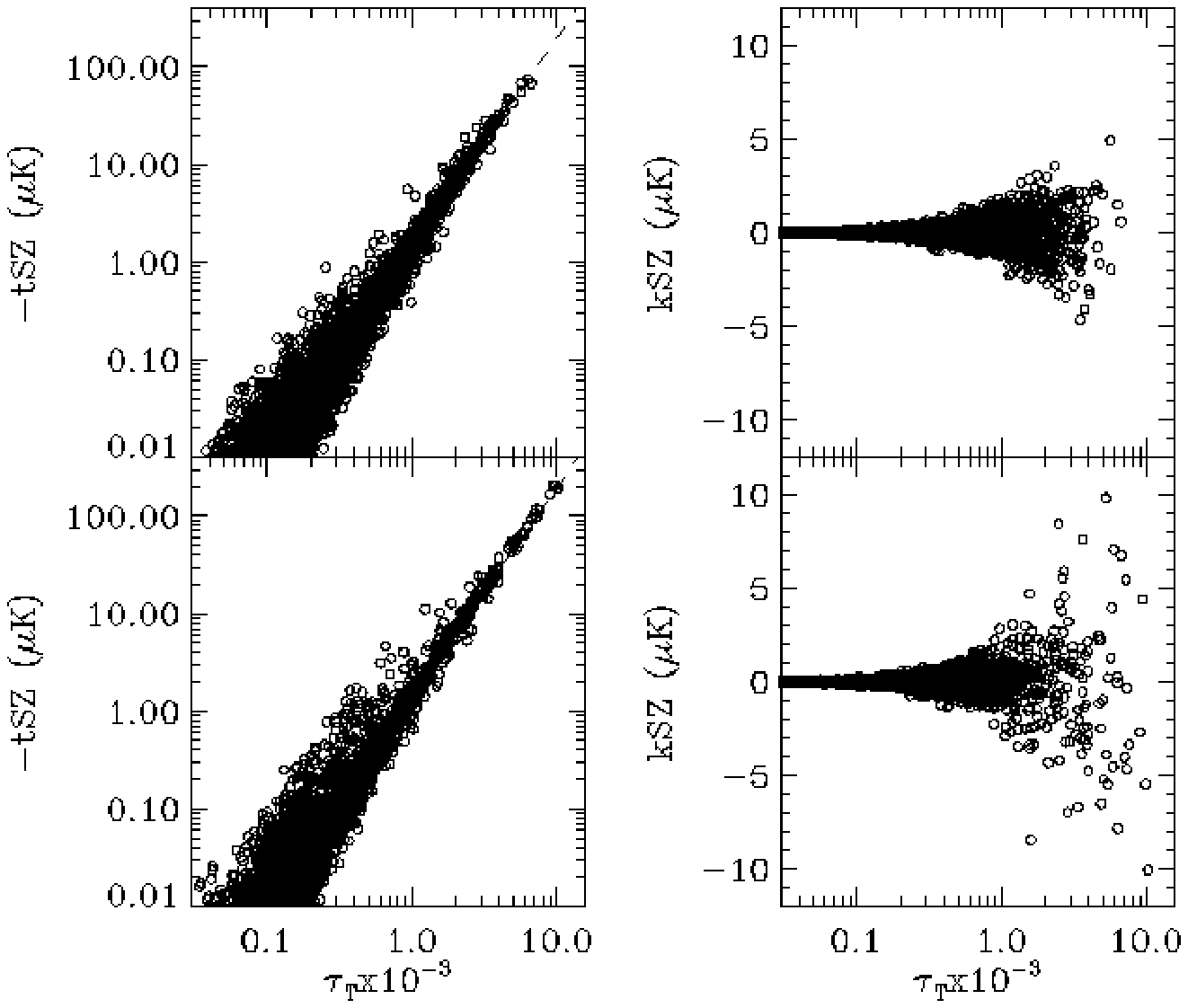}
\caption{Amplitude of the 
tSZ and kSZ fluctuations corresponding to peaks in the OCDM $\tau_T$ map 
at $z=0$. Upper panels show peaks
outside the 45 regions with $\delta_8\ge 2$. Lower panels 
show peaks within these 45 regions.  The dashed lines in the
left panels show that the tight tSZ--$\tau_T$ relation is
well described by the law tSZ($\mu$K$)=-2\times 10^6\tau_T^2$. 
} 
\label{fig:peaks}
\end{figure*}

Figure \ref{fig:maps} shows the kSZ map corresponding to Figure \ref{fig:maptSZ}.
There are many pairs of close fluctuations with opposite sign.
The large scale structure clearly appears as a fuzzy filamentary network
of small spots, which are the dark matter halos constituting the filaments. These
are only a few Mpc long and are substantially curved.
Because of this curvature, filaments have relatively small optical depth and do not
appreciably contribute to the CMB anisotropies with the kSZ effect.
The largest kSZ contributions comes from high density regions (Figure \ref{fig:peaks}).
All the peaks in the $\tau_T$ map which show kSZ fluctuations
larger than $5\mu$K are within regions with $\delta_8\ge 2$.
There are 41 peaks 
with $\tau_T>4\times 10^{-3}$ of which 12 are outside regions
with $\delta_8\ge 2$. 
There are no significant contributions to the kSZ fluctuations 
from peaks with $\tau_T<10^{-3}$. 
Peaks with $\tau_T>10^{-5}$ obey the relation tSZ($\mu$K$)=-2\times 10^6\tau_T^2$.
Note that the bulk flow of the entire simulation box, when included,
may increase the magnitude of the kSZ fluctuations.
For example, in OCDM,
the rms velocity of spheres of diameter $140 h^{-1}$ Mpc is $\sim 90$
km s$^{-1}$, and some kSZ fluctuations in Figure \ref{fig:peaks}
would increase by $\sim15\%$.

Figure \ref{fig:velzooms} shows a supercluster region.
Because the clusters are approaching each other, the 
kSZ peaks have opposite signs: this is the characteristic signature
of the velocity field within superclusters. 
The maps shown 
in Figure \ref{fig:velzooms} are 22.2 arcmin on a side and 
have resolution 0.4 arcmin; the two largest tSZ decrements have
separation of 5 arcmin. 
Examples of neighboring positive and
negative kSZ peaks are also shown in Figure \ref{fig:models} for a supercluster
at $z=0$ as it would appear in our three cosmological models. The angular
dimension of these maps depends on the supercluster distance.
The $\tau$CDM, $\Lambda$CDM and OCDM models yield fluctuations 
with increasing magnitude.
A full comparison of the magnitude of the SZ effects in our cosmological models 
will be discussed elsewhere. Here, we emphasize that, in all models,
superclusters present neighboring tSZ decrements 
and kSZ fluctuations with opposite sign.
Therefore our supercluster identification method is independent of the cosmology.

Table 1 lists  the properties of 
the two largest tSZ decrements in 
a few dense superclusters at different redshifts in the OCDM box. 
In some cases, superclusters have a peculiar feature: 
the kSZ fluctuations corresponding
to the two tSZ decrements are not the largest fluctuations in 
the kSZ maps. We have searched for the kSZ fluctuations which are more than twice 
the kSZ fluctuation corresponding to the tSZ decrement. 
We find 9 such shifted kSZ fluctuations, out of the 3 superclusters at 5
different redshifts listed in Table 1. For example, 
in supercluster \#2 at $z=0$, there are two kSZ fluctuations of $-7\mu$K and $6\mu$K at
a projected distance 
$\sim 1h^{-1}$ Mpc and $\sim 3h^{-1}$ Mpc from the tSZ decrement $-172\mu$K, whose
corresponding kSZ fluctuation is $-3\mu$K.
These shifted kSZ peaks originate from clusters which are
different from the clusters yielding the two largest tSZ decrements
and are not necessarily relaxed.

The presence of large kSZ fluctuations and tSZ decrements 
that do not overlap is confirmed by
the rms of the map fluctuations within the supercluster
regions. In Table 1 we list two rms's, including or excluding 
two squares of $2h^{-1} $ Mpc on a side, each centered on one 
of the two largest tSZ decrements. The tSZ rms drops substantially when
these two squares are excluded. However, this is not always the
case for the $\tau_T$ and kSZ rms's. Therefore, within
superclusters, there may be halos with large peculiar velocities and
sufficiently high Thomson optical depth which cause kSZ fluctuations
larger than the fluctuations due to the most massive clusters.
We finally note that in our two low density models, amplitudes
of both the tSZ and kSZ effects do not show any substantial evolution in 
the redshift range $0\le z\lesssim 0.7$.
At higher redshifts, however, the average amplitudes decrease appreciably. 

\section{Conclusion}
\label{discussion}
The recent development in the microwave detector technology, the
planned MAP, Planck Surveyor, and FIRST experiments, large ground arrays (e.g. ALMA and in
Antarctica), and various balloon projects will provide high precision
measurements of CMB temperature fluctuations on small scales. 
Here, we have seen that peculiar motions of clusters in supercluster
environment produce detectable arcminute anisotropies.  These
kSZ fluctuations come in the form of peaks of opposite signs
but they are an order of magniture smaller than temperature decrements
estimated from the tSZ effect in the Rayleigh-Jeans limit. 

In single-band low frequency CMB observations, superclusters at any redshift
can be identified as well separated areas of
neighboring non-gaussian large decrements superimposed on cosmological
and foregrounds contributions which are more uniformly distributed.
Follow-up observations
in the millimeter and submillimeter bands can disentangle the thermal and kinematic
contributions. We find that the largest thermal and 
kinematic fluctuations can originate from
different clusters within the same supercluster and, therefore, might not 
always be perfectly overlapping.

\acknowledgments The GIF $N$-body simulations were carried out at the
Computer Center of the Max-Planck Society in Garching and at the EPPC
in Edinburgh, as part of the Virgo Consortium project.  A. N. is grateful
to the MPA for the hospitality.

\begin{figure*}
\epsscale{0.5}
\caption{Map of the tSZ decrements (black spots)
in the OCDM box at $z=0$. The comoving
resolution is $0.2 h^{-1}$ Mpc.  
To increase the contrast, decrements larger than $3$ times the rms$=1.2\mu$K are saturated.
The 15 most dense regions are located by squares of $8 h^{-1}$ Mpc on a side.
All but one of the high density region pairs, whose squares overlap on the map, are nearby
in 3D: the separations between the two maxima defining each region 
are $<12 h^{-1}$ Mpc. The pair 2-8 only is a projection effect,
with a separation between the maxima of $41h^{-1}$ Mpc.  }
\label{fig:maptSZ}
\end{figure*}

\begin{figure*}
\epsscale{0.5}
\caption{Map of the kSZ fluctuations corresponding 
to Figure \ref{fig:maptSZ}. 
White (black)
spots correspond to positive (negative) fluctuations. To increase
the contrast, fluctuations larger than $3$ times the rms$=0.12\mu$K are saturated.
The 7 most dense superclusters are located by squares of $8 h^{-1}$ Mpc on a side.
Superclusters \# 1 and \# 3, and superclusters \# 6 and \# 5 are nearby in 3D:
the separations between the two maxima defining each supercluster are 9 and 8 
$h^{-1}$ Mpc, respectively.}
\label{fig:maps}
\end{figure*}

\begin{figure*}
\epsscale{0.5}
\vskip -0.5 truein
\caption{OCDM $\tau_T$, tSZ and kSZ maps 
in  supercluster  \# 5 at $z=0.5$. 
Maps have comoving size $8\times 8h^{-2}$ Mpc$^2$ and $0.2h^{-1}$ Mpc comoving resolution.}
\label{fig:velzooms}
\end{figure*}

\begin{figure*}
\epsscale{0.5}
\vskip -0.5 truein
\caption{kSZ maps in the supercluster region \# 4 in our three cosmological models
at $z=0$. These models have the same initial random phases,
so they yield the same large scale matter distribution at $z=0$.
Maps have comoving size $8\times 8h^{-2}$ Mpc$^2$ and $0.2h^{-1}$ Mpc comoving resolution.}
\label{fig:models}
\end{figure*}

\end{document}